# Microdosimetry in ion-beam therapy: studying and comparing outcomes from different detectors

**Giulio Magrin**
EBG MedAustron, Marie Curie-St. 5, 2700 Wiener Neustadt, Austria
(8 March, 2018)

E-mail: giulio.magrin@medaustron.at

**Abstract.** Experimental studies of microdosimetry in therapeutic ion beams have been performed using several detectors. The differences among the microdosimeters lie on the shapes, the site sizes, and the material. Coin-shaped solid-state detectors made of silicon or diamond with thickness varying from 0.3 μm to 10 μm, as well as proportional counters with spherical and cylindrical sensitive volumes filled with tissue-equivalent gas are the microdosimeters used in therapeutic proton and carbon-ion beams. One goal of microdosimetry in the clinical environment is providing a reliable and repeatable specification of the radiation quality of the radiation field. A methodology should be developed to provide, independently from the heterogeneous information collected with the different detectors, a detector-independent specification of the radiation quality. Historically the specification of the radiation quality is provided either, in terms of Linear Energy Transfer (LET) or in terms of lineal energy, $y$. The first part of this study focuses on identifying the correlation between the distributions of LET and the lineal energy spectra as well as the correspondence between their mean values calculated in frequency and in dose. The evaluation is inspired by the method of LET analysis described by Kellerer [Kellerer, 1972] with adaptation to the peculiarities of the ion-beam therapy where the pristine irradiation is unidirectional and made of a single species of ion essentially mono-energetic. The second objective of this study is to interpret the spectra collected by a slab and perform the necessary conversion to estimate what the spectrum would be if it was collected by a detector different in shape, material, or size. An example is provided using as starting point the simulated lineal energy spectrum of carbon ions impinging a slab detector of graphite and using the method to convert it to the spectra that would be obtained with spherical, cylindrical, and slab detectors made of water in the same radiation field.

## 1. Introduction

Radiation quality is defined as the full spectrum of particle types and the energies. The term 'specification' of radiation quality is used since the funding paper on microdosimetry of H.H. Rossi to indicate that microdosimeter outcomes show important characterizing of the radiation, although not strictly defining it [Rossi, 1959]. The same term is also used in ICRU report on "The quality factor in radiation protection" proposing two options for the specification of radiation quality, in terms of LET[1] or in terms of lineal energy.

---

[1] Here and below LET is intended as unrestricted

Moving from radiation protections to radiation therapy the characteristics of the radiation drastically change, and, although there is consensus in indicating that the radiation quality should be specified using distributions and not simple mean values, the question remains on what quantity to choose, LET or lineal energy. For historic reasons LET is frequently used in the medical field so that the most common way of characterizing the quality is referring to it as low-LET and high-LET radiation. Using lineal energy is not common and this is due the scarcity of experimental microdosimetric data and on the dependence of the results on the detector characteristics. The scenario is gradually changing and several researchers published microdosimetric data collected with a variety of detectors in therapeutic proton and carbon-ion beams. It is important to consider that there is no ideal detector for ion-beam therapy since no one is optimal to cope at the same time with all requirements: high sensitivity, small cross section, good resemblance to biological target, and clinical compatibility.

It is important to separate the different ways in which microdosimetry is employed in the framework of ion-beam therapy which can be essentially distinguished in two cases, first the investigation of the effect of the energy deposition in cells and animals to be used to study and optimize the radiation in *preparation* of the therapy, and second the routine analysis of the beam characteristics to test the *execution* of the treatment and be the base for retrospective studies. This work pursue the second, recognizing the need, after the recent developments, to focus in a theoretical approach and to identify the key parameters to describe radiation quality for the treatment.

This work, focusing on the so-called condition of 'method of LET analysis proposed by Kellerer, studies the energy imparted by therapeutic ion beams to detectors which are different in shapes and materials. The emphasis is in particular in three conditions which are uncommon for the radiation protection environment and standard Tissue Equivalent Proportional Counters (TEPC) microdosimeters: charge particles impinging with a unique direction, detectors of non-tissue equivalent material, and sensitive volumes shaped as slabs. The expected outcome is to obtain a characterization of the radiation which is independent on the detector types so to be able to specify univocally the radiation quality.

## 2. Background
### 2.1. Bases of microdosimetry

Microdosimetry is a mature field of research and in the past 60 years many publications and monographs described the approach and the fundamental parameters. Here are omitted all definitions of the microdosimetric quantities, their correlations, and the way in which the spectral representations are obtained. Publications as ICRU reports on 'Fundamental quantities and units in ionizing radiation' [ICRU, 2011] and on report 'Microdosimetry' [ICRU, 1983] are the reference for the reader interested in elaborating on concepts as energy deposited by a single event $\epsilon_1$ (in the text it will simply indicated as $\epsilon$), the lineal energy, $y$, its values averaged in frequency, $y_F$, and in dose, $y_D$, the cumulative distributions of lineal energy, $F(y)$ and $D(y)$, the density distributions of lineal energy $f(y)$ and $d(y)$, and LET.

### 2.2. The relative variances of the lineal energy

The experimental determination of the pulse amplitude, corresponding to energy imparted on the detector site by a single event of energy deposition, is affected by several factors. A general way to take into account these factors is to describe their individual contribution to the relative variance associated to $\epsilon$. The distribution of the chord length –the cumulative distribution of a chord length is represented here as $X(\ell)$ for a generic detector shape while the corresponding density distribution is represented by $x(\ell)$, the uppercase lowercase convention is maintained thought the paper to indicate cumulative or density distributions of any variables– accounts for the differences of path of the particle within the volume and is related exclusively to the detector shape and the irradiation direction. Its variance is indicated as $V_C$. The energy-loss straggling –related to the number of ion-electron collisions taking place in the site and on to the large difference of the energies exchanged in a single collision– is associated to the variance of energy loss straggling $V_S$. The δ-ray escape –δ-rays, the electrons resulting in the collisions, can have a range large enough to escape the detectors site– is associated to the variance $V_δ$. Experimental factors –which include non-linearity of electronic readout chain and Fano factor– are associated to the variance $V_{exp}$. Gas multiplication, which contributed to $V_{exp}$ in TEPC, is obviously not present in solid-state detectors, nevertheless local lattice irregularities and charge traps can have an effect on the amplitude of the signal and, consequently, contribute to experimental variance.

The last factors to consider are relevant at the lowest energies and relate to the curvature of the track and to the fact that detectors are not dimensionless so that the particles may have significant changes in LET when traversing them or may stop inside. These are named below 'range factors'.

Each factor discussed before depends on the radiation conditions and on the detector characteristics. An important analysis concerns the restrictive conditions for which the total relative variance associated to $\epsilon$ can be expressed exclusively by $V_L$ and $V_C$, while $V_S$, $V_δ$, $V_{exp}$, and range factors are negligible. This was discussed by Rossi [Rossi et al., 1955] and Kellerer [Kellerer et al., 1975] under the so-called "method of LET analysis". In this case the total variance is approximated by:

$$V \cong V_C + V_L \qquad (1)$$

Kellerer identified the condition for which (1) is fulfilled considering on two parameters, the energy of the particles forming the isotropic radiation and the size of a spherical detector made of water. The diagram of Fig. 1 (adapted from Kellerer's for carbon ions) displays the areas where different effects are more relevant. The ions collected under the conditions of region *I* have range –estimated using SRIM tables– equal to or lower than six times the sphere diameter. Region *III* corresponds to the condition in which more than 10% of the δ-ray energy is deposited outside the detector. Region *IV* indicates the condition in which the variance of the energy loss straggling exceeds the variance of the chord length which, for spheres is 1/8 (see section 3.7.4 for further details).

Unlike the general case of completely unknown radiation, in therapeutic pristine ion beams is well defined in energy and composed by a single element. The particles of the beam throughout their path in the target have small deviations from the mean energy and are still composed, in majority, of the same ion species. The energies of the particles in a single beam are represented in the diagram by a segment parallel to the *x* axis whose amplitude is small since it represents the energy spread of the particles. With an appropriate choice of the detector size at a specific depth, this segment can be completely included in region *II* of the diagram.

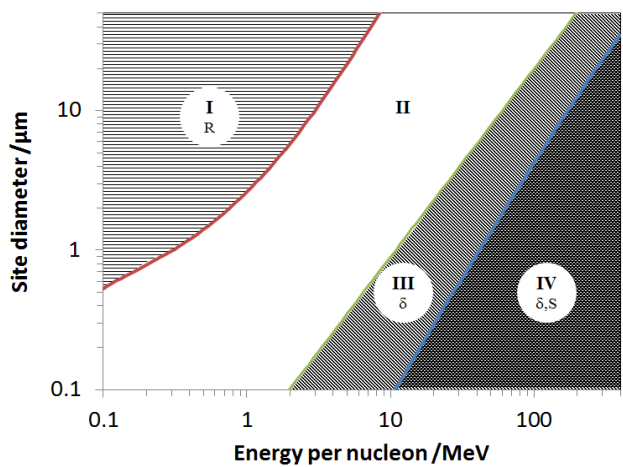

Figure 1. Diagram of the relevance of the range factors, the δ-ray escape, and the energy loss straggling as function of the carbon-ion energy per nucleon and of the diameter of the spherical water microdosimeter. See text for description of the four different regions.

Equation 1 is valid in region *II* and consequently the energy imparted $\epsilon$ can be represented, with good approximation, by the product of the random variables which expresses the chord length of irradiated site, $\ell$, and the LET of the particle, $L$:

$$\epsilon \cong \ell \cdot L \qquad (2)$$

Diagrams similar to that one represented in Fig. 1 for spherical detectors can be obtained for cylindrical and slab detectors. In a new sub-region of region *II*, named *II-intersection*, resulting from the intersection of the regions *II-sphere, II-cylinder,* and *II-slab* obtained for the three different shapes, Eq. 2 is valid independently on the detector shape. It is reasonable to assume that *II-intersection* is not substantially smaller than the region *II* represented in Fig. 1. It is very convenient to use region *II-intersection* for the comparison of the performance on the different microdosmeters since the experimental results collected with the microdosimeters will differ only for the contribution of the chord-length distribution, $X(\ell)$, being the LET distribution the same for all detectors and all other variances negligible. This is the starting point of the analysis described in section 3.1.

The suggestion on how to obtain the diagrams for non-spherical shape and for unidirectional radiation is carried out in section 3.7, where 3.7.1 and 3.7.2 focus on incomplete crossing and large LET variations, 3.7.3 on escape of δ rays, and 3.7.4 on energy loss straggling.

*2.3. Variance contributions in therapeutic ion beams.*

Proton and carbon-ion beams release the highest energy at the end of their path forming the so-called Bragg peaks. In order to conform the dose to the tumor target, the energy of the ion beams is modulated and the spread-out Bragg peak, SOBP, is obtained.

For therapeutic ion beams, the relevance of each variance component described before changes with the penetration depth. Three distinguished regions can be considered, the entrance, in a generic point of the SOBP, and at the distal part of the SOBP. Their characteristics are described hereafter.

Before entering the target the pristine radiation field is composed by a unique ion species and all its ion have, essentially, the same energy. Consequently $V_L$ is small and the LET distribution is a narrow peak. The spectrum of imparted energies is the result of contributions of the chord length variation and, since the energy is high, of the energy-loss straggling, and of the δ-ray escape. Using a larger detector with size of 10 μm would be convenient in this case to move the 'working point' in the diagram of Fig. 1 away from the region *III* and *IV* and toward region *II*.

In the SOBP, after crossing several centimeters of tissue –in therapy the irradiation depth can be as high as 25 centimeters of water– the particles are more differentiated in energy, the presence of ion fragments increases, and consequently and $V_L$ enlarges. At the same time the beam energy is lower than at the entrance and so is the relevance of the energy loss straggling and δ-ray escape. The condition represented in region *II* of the diagram of Fig. 1 can always be reached providing detectors appropriately sized (few micrometers) to minimize from one side the delta-ray escape and from the other side the range factors.

At the lowest energies there is an increases probability for particles to stop within the detector as well as the probability of significant changes of LET. The parameter known as range straggling, which describes the statistical fluctuations of the ranges for the particle of certain energy in a specific material, is used (see sections 3.7.1 and 3.7.2) to study the percentage of particles stopping in the detector and to assess when the variation of LET in the transversal exceeds a preset value. The optimal detector should have the smallest feasible size, one micrometer or less. Since the microdosimeter size cannot be decreased indefinitely a minimum energy is reached at which the working point in the diagram of in Fig. 1 is not maintained in region *II* and falls in region *I*.

## 3. The Method
### 3.1. Assumptions for therapeutic ion beams

The special restrictions which allow expressing the energy imparted as indicated in Eq. 2 can be described for therapeutic ion beam by the two following conditions: (i) the variation of the LET, while the particle is crossing the detector, is negligible; (ii) the fraction of secondary electrons escaping the sensitive volumes is negligible. Condition (i) is sufficient to guarantee that all range effects are not significant since, as it is described in sections 3.7.2, the number of particles with incomplete transversal is always smaller than number of particles with high LET variation during the transversal. Stating that condition (ii) is satisfied also implies that the straggling of energy loss is negligible since, as shown in Fig. 1, δ-ray escape always exceeds the distortion of straggling of energy loss. For proton and carbon-ion beams and for detector sizes ranging from 1 μm to 10 μm, the assumptions (i) and (ii) are satisfied for particle energies between 1 MeV/u to 80 MeV/u.

To emphasize the specificity of therapeutic ion beams, the further assumption should be considered that: (iii) the particle tracks in the detector are straight and parallel. The near parallelism of the tracks is confirmed, macroscopically, by dosimetric analysis of the therapeutic ion beams in phantoms, and is the fundamental characteristics that allows the conformation of the dose in deep-seated tumors and, ultimately, justifies the use of ion beams in therapy. The strait-line approximation, and tracks parallelism are discussed in section 3.7.5.

In Eq. 2 the energy imparted $\epsilon$ is expressed as the product of the independent variables $\ell$ and $L$, to which the cumulative distributions $X(\ell)$ and $T(L)$, can be associated, respectively. The density distribution of LET, indicated as $t(L)$, is the derivative of $T(L)$ respect to $L$. The cumulative distribution of $\epsilon$ is indicated as $F(\epsilon)$, and its derivative respect to $\epsilon$ is the density distribution of $\epsilon$, $f(\epsilon)$.

The distributions $F(\epsilon)$ and $f(\epsilon)$, can be expressed [Kellerer, 1972] in terms of the distributions of $\ell$ and $L$ as follows:

$$F(\epsilon) = \int_0^{+\infty} t(L) \cdot X(\ell) dL = \int_0^{+\infty} t(L) \cdot X\left(\frac{\epsilon}{L}\right) dL$$

$$f(\epsilon) = \int_0^{+\infty} t(L) \cdot x(\ell) \cdot \frac{1}{L} dL = \int_0^{+\infty} t(L) \cdot x\left(\frac{\epsilon}{L}\right) \cdot \frac{1}{L} dL$$

(3)

Unidirectional radiation and isotropic radiation relate to different chord distributions –except for the case of sphere– which in turn result, according to Eq. 3, to different microdosimetry spectra and different values of the dose-mean lineal energy, $y_D$. For cylindrical and slab sites and unidirectional ion tracks, the distribution of the chord length can be expressed in simple analytical forms, unlike the case of isotropic irradiation.

Chord-length cumulative distribution and probability density distributions are described in Appendix A for spherical and cylindrical shapes of the sensitive volume.

### 3.2. Slab sensitive volume

A further class of detectors is composed by flat detectors which have sensitive volumes shaped as cylinders or parallelepipeds and are irradiated perpendicularly to the base. Both, solid-state and gaseous detectors have been developed with those geometries, having, in general, transversal sizes much larger of the thickness so to assume the shape of a coin (the cylinder) or of a slab (the parallelepiped). For a uniform and unidirectional irradiation that hits the slab detector perpendicularly to the entrance face, the cumulative distribution $S(\ell)$ and the probability density distribution $s(\ell)$ are represented by the unit step functions, and the Dirac function, δ, respectively:

$$S(\ell) = \begin{cases} 0, & 0 \leq \ell < t \\ 1, & \ell \geq t \end{cases} \tag{4}$$

$$s(\ell) = \delta(\ell - t) \tag{5}$$

with, $t$, indicating the thickness of the detector.

*3.3. First and second moment chord length, and maximum chord length,*

The first moment chord length or mean chord length, $\bar{\ell}$, is the mean value of the tracks of the particles within the sensitive volume of the detector under straight-line approximation. The mean-chord length depends on the direction of the radiation so that, for the same detector, isotropic and unidirectional irradiations relate to different values of mean-chord lengths (except for spheres). The probability density functions of the chords length are used to assess the value of $\bar{\ell}$ for different detectors. Indicating with $x(\ell)$ the chord-length density function for a generic detector shape, the mean chord value is obtained from:

$$\bar{\ell} = \int_0^{+\infty} \ell \cdot x(\ell) \, d\ell \tag{6}$$

The second moment of the chord length, $\overline{\ell^2}$, is also calculated using the density distributions of the chords:

$$\overline{\ell^2} = \int_0^{\infty} \ell^2 \cdot x(\ell) \, d\ell \tag{7}$$

The maximum chord length, $\ell_{max}$, is the maximum particle track under the straight-line approximation and depends as well on the directional characteristics of the radiation.

For complex shapes and irradiation conditions, for which the chord-length density distribution is not described by a simple analytical function, the values of $\bar{\ell}$, $\overline{\ell^2}$, and $\ell_{max}$ can be obtained by simulation. The numerical correlations between $\bar{\ell}$, $\overline{\ell^2}$, and $\ell_{max}$ for the spherical and cylindrical shapes are indicated in Appendix A. For irradiations perpendicular to the base of cylinders or slabs, $\ell_{max}$ corresponds to their thickness, $t$.

For a slab detector the mean value of the chord, $\bar{\ell}$, is (see Eq. 5 and 6):

$$\bar{\ell} = \int_0^{+\infty} \ell \cdot s(\ell) \, d\ell = \int_0^{+\infty} \ell \cdot \delta(\ell - t) \, d\ell = t = \ell_{max} \tag{8}$$

and the second moment of the chord, $\overline{\ell^2}$, is (see Eq. 5 and 7):

$$\overline{\ell^2} = \int_0^{+\infty} \ell^2 \cdot \delta(\ell - t) \, d\ell = t^2 \tag{9}$$

*3.4. Edge values and self-calibration*

The edge in the spectrum is formed by those particles which impart the maximum energy to the detector site as the result of two concurring conditions, the particles have the energy which corresponds to the maximum LET for the ion species and for the specific material, $L_{max}$, and they transverse the detector trough the maximum chord.

The values of $\bar{\ell}$ and $\ell_{max}$ are used, in experimental microdosimetry, for estimating the lineal energy value of the edge of a specific ion in the so-called method of self-calibration of microdosimetric spectra [Rossi et al., 1996]. When several ions with electron stopping power values close to $L_{max}$ cross the detector along or near the maximum chord $\ell_{max}$, an edge is formed on the spectrum, and the corresponding value of lineal energy is:

$$y_{edge} = \frac{\ell_{max}}{\bar{\ell}} L_{max} \qquad (10)$$

The value of $L_{max}$ is provided by the lookup tables of electron stopping power as those published by ICRU [ICRU, 1993, 2005], NIST [Berger et al. 2005], SRIM [Ziegler et al., 2010], and others. The electron stopping power values for specific ions and materials correspond to the unrestricted track LET values. It is worth mentioning that, strictly speaking, Eq. 10 is valid only under the assumptions (i) and (ii). Condition (ii) is always satisfied given the very low energy of the ions in the edge region. Condition (i) instead, is valid only in case of very thin detectors as it can be deducted observing the very rapid variation of the electron stopping power with particle range in the low energy region. In the case of carbon ions in diamond at the energy of the maximum stopping power the particle range is approximately 2.8 μm and therefore the detector thickness should not exceed 0.5 μm, according to the rule adopted for Fig. 1. For detector thickness above that value, to avoid systematic overestimations of the edge, the maximum electron stopping power value $L_{max}$ should be substituted with the maximum average of $L$ within the detector thickness. A description of the procedure is described in the following section 3.7.6. An implicit assumption for using the calibration Eq. 10 is that the stopping power values are good approximations of the lineal energies for corresponding particles and material. This assumption is discussed, for different experimental conditions, in the following section 3.5.1.

In order to calibrate the experimental spectrum in lineal energy, the edge value of the lineal energy must be associated to the corresponding edge channel position. An analytical method for assessing the edge channel position from an experimental spectrum has been described by Conte [Conte et al. 2013].

The method of self-calibration is necessary in all circumstances in which the signal generated by the detectors provides a relative measure of the ionization as in the case of detectors using gas multiplications. In solid-state detectors the relation between energy imparted and pulse height is rather stable and the calibration is done knowing the energy deposited by a known radiation. For these detectors the self-calibration method can be used as verification.

### 3.4.1. Lineal energy mean values, $\bar{y}_F$ and $\bar{y}_D$,

From the integral (14) it derives that the frequency and the dose mean lineal energy are expressed in terms of LET and chord length quantities as follows [Kellerer et al. 1975]:

$$\bar{y}_F = \bar{L}_T \qquad (11)$$

$$\bar{y}_D = \bar{L}_D \frac{\overline{\ell^2}}{\bar{\ell}^2} \qquad (12)$$

Equation 11 shows how, independently on the detector shape, the frequency mean lineal energy $\bar{y}_F$ coincides with the track-mean LET.

In the case of a slab detector an analogous correlation holds also for the dose mean values of lineal energy and LET as it results from Eq. 12 using the results of Eq. 8 and 9:

$$\bar{y}_D = \bar{L}_D \cdot \frac{\overline{t^2}}{\bar{t}^2} = \bar{L}_D \qquad (13)$$

The results for spherical and cylindrical detectors are reported in Appendix A.

### 3.4.2. Chord-length and microdosimetric parameters

Table 1 summarizes chord-length parameters for different detector shapes as described in previous sections, in Appendix A, and in literature. The value of the ration $\overline{\ell^2}/\overline{\ell}^{\,2}$ calculated using Eq. 8 and 9 coincides with the ratio of the dose-mean lineal energy, $\bar{y}_D$, and the dose-mean LET, $\bar{L}_D$. The values of the ratio $\ell_{max}/\overline{\ell}$ and $L_{max}$ can be used in Eq. 10 to assess the lineal energy at the spectrum edge, $y_{edge}$. For all detectors, independently on their shapes, the ratio $\bar{y}_F/\bar{L}_F = 1$.

**Table 1**. Chord lengths parameters estimated for different detector shapes for irradiation directions

|  | $\overline{\ell}$ | $\overline{\ell^2}$ | $\ell_{max}$ | $\overline{\ell^2}/\overline{\ell}^{\,2}$ $(=\bar{y}_D/\bar{L}_D)$ | $\ell_{max}/\overline{\ell}$ $(=y_{edge}/L_{max})$ |
|---|---|---|---|---|---|
| **Sphere** diameter d, both, isotropic and non-isotropic radiation | $\frac{2}{3}d$ | $\frac{1}{2}d^2$ | $d$ | $9/8 (= 1.125)$ | $3/2 (= 1.5\,d)$ |
| **Cylinder** diameter $d$, beam axis normal to axis | $\frac{\pi}{4}d$ | $\frac{2}{3}d^2$ | $d$ | $\frac{32}{3\pi^2} (\cong 1.081)$ | $4/\pi\ (\cong 1.27)$ |
| **Cylinder** diameter $d$, isotropic radiation | $\frac{2}{3}d$ | $\cong \frac{5}{9}d^2$ | $\sqrt{2}\,d$ [1] | $\cong 5/4 (= 1.25)$ [1] | $3\sqrt{2}/2\ (\cong 2.12)$ |
| **Slab and cylinder** thickness $t$, beam axis normal to base | $t$ | $t^2$ | $t$ | 1 | 1 |

[1] Extrapolated from Kliauga [Kliauga et al.,1995]

For spherical volumes the results are independent on the directional characteristics of the irradiation and this is due to the isotropic shape of the sphere. At the contrary in the case of cylindrical shapes, the directional characteristics of the radiation are affecting the results, as shown in Tab. 1. So, two cylindrical detectors, one under isotropic radiation the other under unidirectional radiation perpendicular to the axis, have coincident mean chord length when the ratio between their diameters is 1.18. The dose-mean lineal energy value, $\bar{y}_D$, (the value of the lineal energy edge, $y_{edge}$) evaluated for a cylindrical detector considering an isotropic radiation field is 16% (67%) larger than the value for unidirectional field.

### 3.5. Conversion of spectra for different detector shapes

The distributions of single event of energy-deposition coincide, in value, to the distributions of lineal energy when the condition $\epsilon = y \cdot \overline{\ell}$ is fulfilled. Although the domains of the two distributions are different, the same notation of $F$ and $f$ used in Eq. 3 is maintained and the distributions in lineal energy are rewritten:

$$F(y) = \int_0^{+\infty} t(L) \cdot X(\ell)dL = \int_0^{+\infty} t(L) \cdot X\left(\frac{y \cdot \overline{\ell}}{L}\right)dL \tag{14}$$

$$f(y) = \int_0^{+\infty} t(L) \cdot x\left(\frac{y \cdot \overline{\ell}}{L}\right) \cdot \frac{1}{L}dL \tag{15}$$

The conversion from a spectrum collected from a detector of a certain shape to a detector of a different shape can be done considering the method described by Kellerer [Kellerer, 1972]. Starting from Eq. 14, and knowing the experimental spectrum of the lineal energy obtained with a general detector shape, $F(y)$, for which the distribution of the chord length $X(\ell)$ is known, Kellerer evaluated numerically the LET distribution $t(L)$. The method, which uses and the Fourier transform, introduces an approximation since it foresees to convert the distributions to a logarithmic scale. In the case of experimental spectra obtained with slab detectors the procedure is strongly simplified and does not requires the logarithmic representation as it is described in the following section.

*3.5.1. Spectra of lineal energy distribution of lineal energy from slab to generic shape.*

Let us consider the case of slab detectors under the assumptions (i), (ii), and (iii) in which the particles are imping perpendicularly to the detector face and indicate with the star '*' the values and the distributions of lineal energy for the slab detector.

It is straightforward to recognize that, when the numerical values of $y^*$ and $L$ are coincident, the two cumulative distributions $F^*(y^*)$ and $T(L)$ are identical.

A formal verification can be done considering Eq. 14, choosing a value $y_A^*$ within the interval in which $F$ is defined, and using the chord distribution $S(\ell)$ of Eq. 4:

$$F^*(y_A^*) = \int_0^{+\infty} t(L) \cdot S\left(\frac{y_A^* \cdot t}{L}\right) dL \tag{16}$$

From its definition, $S(y_A^* \cdot t/L) = 0$, for $(y_A^* < L)$, so the integral in Eq. 16 becomes:

$$F^*(y_A^*) = \int_0^{L_A = y_A^*} t(L)\, dL \overset{\text{def}}{=} \int_0^{L_A = y_A^*} \frac{dT(L)}{dL}\, dL = T(L_A) \tag{17}$$

where $L_A$ is the value of the LET numerically coincident with $y_A^*$.

Since Eq. 17 is valid for any value of the interval in which $F$ is defined, the correlation between the distributions is general:

$$F^*(y^*) = T(L) ; \tag{18}$$

Analogous result is found for the density distributions:

$$f^*(y^*) = t(L) \tag{19}$$

The cumulative distribution $F(y)$ that would be collected with a detector of generic shape, whose chord-length cumulative distribution $X(\ell)$ is known, can be obtained substituting $f^*(y^*)$ to $t(L)$ in Eq. 14:

$$F(y) = \int_0^{+\infty} f^*(y^*) \cdot X\left(\frac{y \cdot \overline{\ell}}{y^*}\right) dy^* \tag{20}$$

The lineal energy spectra are collected experimentally and are represented by discrete distributions. The element of the discrete distribution is represented as:

$$F(y_j) = \sum_i f^*(y_i^*) \cdot X\left(\overline{\ell}\frac{y_j}{y_i^*}\right) \tag{21}$$

and, correspondingly, the discrete density function becomes:

$$f(y_j) = \sum_i \frac{f^*(y_i^*)}{y_i} \cdot x\left(\overline{\ell}\frac{y_j}{y_i^*}\right) \tag{22}$$

If, representing the spectra, the jumps between adjacent bins are identical in the slab and the generic detector, Eq. 21 becomes:

$$F(y_j) = \sum_i f^*(y_i^*) \cdot X\left(\overline{\ell}\frac{j}{i}\right) \qquad (23)$$

In the case of spherical and cylindrical detectors the distributions of the chord lengths $X(\ell)$ and $x(\ell)$ to be used in Eq. 21 and Eq. 23 are shown in Appendix A.

Representing the density distribution of lineal energy using Eq. 22 may lead to distorted results. In particular cases the discrete chord-length distribution has significant changes between two consecutive bins. For instance in cylinders the chord density distribution $c(\ell)$ (see Eq. A.4 in Appendix A) is divergent when $\ell$ approaches $d$ and therefore it is very sensitive to the choice of the bin amplitude, easily distorting the results. A consequence is that the sum of the discrete density function $\sum_i x(\overline{\ell} \cdot y_j / y_i^*)$ differs from unity and it is not an accurate indication of the probability.

When this happens the derivative $x(\overline{\ell} \cdot y_j / y_i^*)$ in Eq. 22 should be approximated by the incremental ratio:

$$x(\overline{\ell} \cdot y_j / y_i^*) \cong \frac{X(\overline{\ell} \cdot y_{j+1}/y_i^*) - X(\overline{\ell} \cdot y_j/y_i^*)}{(y_{j+1} - y_j) \cdot \overline{\ell}/y_i^*} \qquad (24)$$

which preserves the normalization.

### 3.6. Conversion to spectra of different detector materials

The capability of converting spectra collected with detectors of one material to the spectra that would be collected in a different material would allow comparing heterogeneous historical data on ion beams and to refer the microdosimetry outcomes (the lineal energy distributions and mean values) to a unique clinical standard material. Today there is no defined standard material and the choice is still open. One possibility is to select the tissue-equivalent (TE) materials, generally used in traditional microdosimetry, as the propane-based TE gas. In this study the choice is, instead, to refer the microdosimetric parameters to water, considering its relevance in radiation therapy for what concerns both, dosimetry and treatment planning purposes.

#### 3.6.1. Single value conversion

A method normally used to adjust the frequency distributions obtained in one material to a different material is to 'recalibrate' the spectrum for the new material based on a single conversion factor. This process is done assigning to a point which can be easily identified in the experimental spectrum –as the ion edge or the spectrum peak obtained for a known radiation field– the corresponding value estimated for the new material. Alternatively the conversion factor can be estimated averaging the ratio of the electron stopping power values for the two materials over a range of energies. In both cases the conversion factor is deducted from published lookup tables or from previous experimental data. The recalibrated spectrum is obtained rescaling linearly all other bin values.

This method is applicable if the values of the mass electron stopping power of the two materials are proportional (or very close) for all energy values and for all particles, primary ions and fragments. This condition is fulfilled for instance when a carbon-ion spectrum in T.E. propane-based gas is converted to the spectrum in water. The correlation between the electron stopping power of the two materials is well described by a linear function with deviations non exceeding 5% in the range of energy of ion-beam therapy.

### 3.6.2. Conversion from graphite to water

In the case of silicon and diamond detector, the conversion to water cannot be achieved using, for the full range of clinical energies, a single scaling factor. For those materials, the electron stopping power ratio to water varies substantially at different energies, with deviations up to 40% for silicon and up to 20% in graphite (graphite is used in lookup tables in place of diamond), as it is shown in Fig. 2 using the stopping power values from SRIM look tables. Also limiting the energy interval to the range between 5 MeV and 60 MeV, the differences are still large, 36% in silicon and 16% in graphite. This means that, to avoid misrepresentation of a spectrum converted to water, the recalibration method should be based on different conversion factors each one defined for a small interval of particle energy.

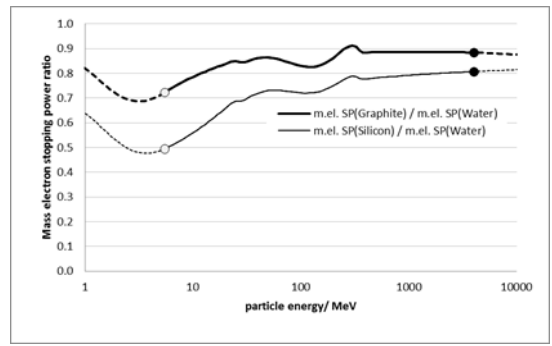

Figure 2. Thick line: Ratio of the carbon-ion mass electron stopping power in graphite and water as function of the particle energy. Thin line: Ratio of the carbon-ion mass electron stopping power in silicon and water as function of the energy. The data are obtained with SRIM. The solid lines indicate the energy intervals meaningful for carbon-ion beam therapy purposes: the upper limits (filled circle markers) coincide with the maximum carbon-ion energy typically employed in therapy, 4800 MeV; the lower limits (empty circle markers) indicate the energy value corresponding to the maximum stopping power.

Unrestricted LET values alternatively called 'linear electron stopping power' [ICRU, 1998], are available for a variety of ions and targets as function of ion energy and range. One may neglect, for the moment, the fragmentation of the carbon ions. The electron stopping power as function of the energy is a monotone decreasing function, for different material, from values above 5 MeV (4 MeV for water and 5.25 MeV for graphite). For carbon-ion energy up to 150 MeV (12.5 MeV/u) the assumptions (i) and (ii) discussed before are fulfilled for microdosimeters of about 1 µm and, as discussed in section 3.5.1, the lineal energy spectrum collected with an slab detector $f^*(y^*)$ is a good approximations of the LET distribution $t(L)$ and the two curves overlap when the values of $y$ and $L$ are represented in the same scale.

From the monotonicity of the electron stopping power as function of the energy descends that one value of LET corresponds to a unique value of energy. It is then possible to correlate the lineal energy values of the experimental spectra and the energy of the particles that produced them. Vice versa, knowing the energy values of each particle it is possible to deduct the corresponding LET value in a different material and, by reiteration, foreseen the full LET distribution in the new material. This distribution also coincides with the microdosimetric spectrum in lineal energy that would be collected by a slab detector made of the new material.

Operatively the material conversion of the spectra of carbon ions from diamond to water is done following the steps described hereafter. Lacking specific tables, diamond mass electron stopping power vales are assumed to coincide with those for graphite.

- The experimental discrete spectrum is obtained through a multi-channel analyzer: the abscissa is divided in equally spaced bins of increasing lineal energies, and the ordinate indicates the relative number of counts collected in each bin.

- The value of the lineal energy of each bin is used to compute the corresponding energy of the carbon ions by interpolation in the lookup tables from carbon-ion mass-electron-stopping-power in graphite.
- This energy is used to assess the electron stopping power values in water via lookup tables. This step and the previous one can be reduced to a single step in which the mass electron stopping power tables of graphite and waters, for identical energies, are joint and the value of mass electron stopping power of water is assessed directly via interpolation from the value of the mass electron stopping power of graphite.
- The relative counts of the original bin are assigned to the new bin in water.
- Finally the process is repeated for all original bins to reconstruct a full spectrum in water. It must be noticed that the channels in the new spectrum are not equally spaced. To preserve the representation in channels with the original lineal energy values, the new spectrum should be re-binned and renormalized.

There are some uncertainties arising from this re-calibration process. First, the method assumes that only primary particles are involved although the incidence of nuclear fragments is non-negligible in particular at the end of the range. This issue is discussed on the next section 3.6.3. Second, the energy corresponding to the maximum electron stopping power is not the same for different material. As cited before, carbon ions in water have maximum electron stopping power at 4 MeV and in graphite at 5.25 MeV. Consequently, all the particles impinging the diamond with energy below 5.25 MeV are erroneously allocated in the new water spectrum. This results in a slight drift of the carbon edge in water to lower values.

### 3.6.3. Ion fragmentation and material conversions

The conversion of the spectra for different materials discussed in the previous section assumes that the fragmentation of the primary particle is negligible. In a realistic condition the relative contribution of fragments increases with the depth and high particle heterogeneity is expected getting closer to the Bragg peak.

The presence of fragments is not a large source of distortions in all material conversion as discussed before for conversions between T.E. gas. The following example considers a more general case in which carbon ions and their fragments are impinging the graphite detector. The ratio of mass stopping powers in graphite and water of all carbon fragments, hydrogen, helium, lithium, beryllium, and boron is evaluated using on SRIM data and represented in Fig. 3.a. In this case it is evident that adopting a single conversion curve for all ions may introduce spectra distortions. The differences are as high as 15% occurring at energies where the electron stopping power reaches its maximum and they are more relevant for the heaviest ions (carbon, boron, and beryllium). In those conditions the ions have the energy of few mega electronvolts and limited range (few micrometers in range).

At all irradiation depths the fragmented ions are lower in number compared to the primary particles and the energy imparted by the fragments is sensibly lower than the energy imparted by the primaries. The microdosimetric *frequency* spectra may show visible distortions mainly due to fragmented protons. The distortions are much less evident in the microdosimetric *dose* spectra at least when the detectors are placed at depths within the Bragg curve and not in the tail formed by the fragments afterward.

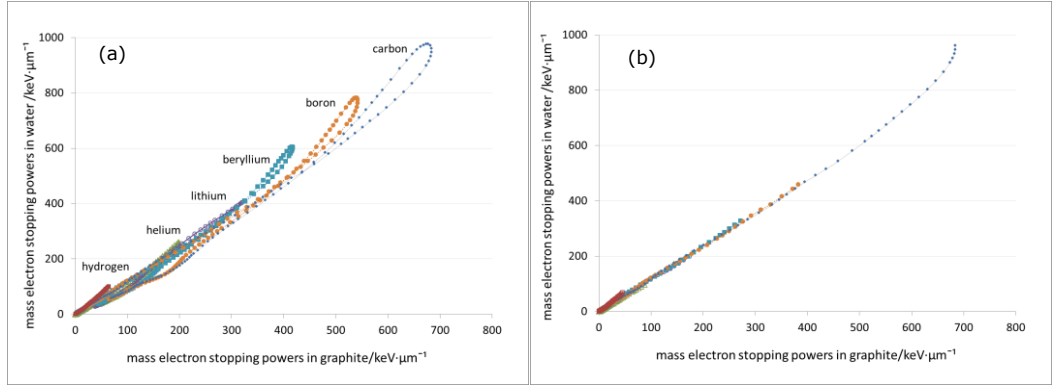

Figure 3. Mass electron stopping power in water vs. mass electron stopping power in graphite for primary ions (carbon) and fragments (boron, beryllium, lithium, helium, and hydrogen) as provided by SRIM: (a) Particles with energy below 1000 MeV; (b) As in (a) but excluding all particles with range below 40 µm.

In Fig. 3.b all stopping powers are re-evaluated excluding the values that correspond to projected ranges in graphite of less than 40 µm. With this cutoff condition the data appear to be well represented by a single curve. Analogous results can be found for silicon based detectors.

From the previous observation it can be seen that if a strategy to univocally identify the low range particles in graphite is put in place, it is also possible to obtain accurate representations of the spectra in water. A possible way to set a range threshold is by using the detectors called 'radiation telescopes'. The telescopes are formed by two independent detectors placed one on top of the other: a thinner '$\Delta E$ layer', with the characteristics of a microdosimeter, and a thicker '$E$ layer' in which low energy particles stop. A single particle crossing both layers produces two simultaneous signals, which combined provide for the same particle LET and total energy, parameters that can identify the particle species. A specific telescope design can be developed to use the $E$ layer to assess if the particle range exceeds or not 40 µm. With it, the spectra conversion from graphite to water is obtained recombining the results of the high and low range particle.

### 3.7. Quantitative assessment of relevant parameters and complementary considerations

#### 3.7.1. Finite range of primary particles

The probability of the particles to stop inside the detector is linked to the position of the detector in relation to the range straggling, which is a collective characteristic of the beam. When a monochromatic ion beam transverse a volume of uniform material, the position in which the particles stop is well-described by a Gaussian distribution [Chu et al., 1995] with standard deviation represented by a function which depends on the mean range, $r$, and on atomic mass, $A$, of the ions as follows:

$$\sigma_r = 0.012 \cdot r^{0.951} \sqrt{\frac{1}{A}} \qquad (25)$$

For protons the standard deviation corresponds to approximately 1.0 % of mean range (the FWHM is 2.3%) and, for carbon ions, to 0.31% of the range (the FWHM is 0.73 %). One may call the region which extends for the FWHM around the mean range value the 'range straggling FWHM'. For clinical purposes, the range straggling FWHM is maintained, for every irradiation condition, at 1 millimeter or more to guarantee dose homogeneity. For superficial irradiations in which the intrinsic range straggling would be too narrow, passive elements, as ripple filters and absorbers, are interposed in the beam lines so to increase the range straggling FWHM.

The probability for a particle to stop in a detector placed in position $x_1$ is described by the following:

$$P(x_1, t) = \frac{1}{\sqrt{2\pi} \cdot \sigma_r} \int_{x_1}^{x_1+t} e^{-\frac{(x-r)^2}{2\sigma_r^2}} dx \tag{26}$$

The ratio, $n_r$, defined as the relative number of particles stopping in the detector out of all particles entering the detector is given by the following:

$$n_r(x_1, t) = \frac{\int_{x_1}^{x_1+t} e^{-\frac{(x-r)^2}{2\sigma_r^2}} dx}{\sqrt{2\pi} \cdot \sigma_r - \int_{-\infty}^{x_1} e^{-\frac{(x-r)^2}{2\sigma_r^2}} dx} \tag{27}$$

The value of $n_r$ increases constantly with the depth $x_1$, while the number of particles reaching the detector progressively decreases. When the detector is placed at depths shorter than the mean range $r$, only few particles are stopping inside and then there is minimal or no spectra distortion. At the depth corresponding to the range, $r$, for a range straggling FWHM of 1 mm, only 0.4 % of the particles stop in a 1µm-thick detector as estimated using Eq. 27. Moving the detector deeper, to the distal part of the range straggling FWHM, the relative number of particles stopping in the detector increases to 1.3% but the total number of particles detected drops to only 0.9% of the original. The detector cross sections adapted to clinical irradiation fluences are of the order of few hundredths of square millimeters and for these conditions the number of events collected in that position are too small to provide a significant spectrum.

### 3.7.2. LET changes within the detector

According to Kellerer's analysis used in section 2.2, the LET variation during the detector transversal is below 10% if the detector thickness is smaller than 6 times the residual range. This constraint is discussed here for therapeutic ion beams estimating, in some critical range conditions, the relative number of particles that exceed the limit of 10% in LET variation.

The evaluation is based on the observation that the absolute variation of LET of the particles having energy equal to or lower than a certain value, can be expressed as monotonic decreasing function of the energy. Also the variation of LET values traversing a detector of thickness $t$, increases with the decrease of the residual range. Because of the monotonic characteristics, evaluating the probability of a particle to experiences a variation of LET exceeding a chosen value (arbitrarily defined as 10%), is analogous to evaluate the probability for the same particle to have an energy below $E'$, estimated from lookup tables as the energy value at which the LET variation is 10%, or to stop within the thickness $t'$ (i.e. the particle mean range at the energy $E'$).

The following example shows the procedure. An ideal water slab microdosimeter of thickness $t = 1$ µm is placed in a carbon-ion beam with 14 centimeters of range and range straggling FWHM of 1 mm (see Eq. 29). The energy for which the electron stopping power variation equals 10% is evaluated to be of 2.75 MeV using the lookup tables. At this energy, the carbon ion has a residual projected range in water of, $t' = 5.1$ µm. It should be noticed that Kellerer's rule of a range to be 6 times the detector thickness is well in agreement with the numerical values of $t'$ and $t$. The probability to have a LET variation exceeding 10% corresponds also to the probability for the particle to stop within a detector of thickness $t'$ and is obtained substituting $t'$ to $t$ in Eq. 26. Analogously, the fraction of particles exceeding 10% of LET variation is obtained substituting $t'$ to $t$ in Eq. 27. In this example, when the detector is placed at the depth corresponding to the mean range, $r$, the value on $n_r(x_1 = r, \ t' = 5.1) = 0.9\%$ and when it is placed at the distal part of the range straggling FWHM is 1.9%.

### 3.7.3. Non-negligible variance of δ-ray escape

In order to assess the effect of δ-ray escape, a simple approach was described for spherical detectors in section 2.2 and illustrated in Fig. 1. For non-spherical detectors the δ-ray escape should be evaluated taking into consideration size and shape of the microdosimeter.

In particular for slab detectors with large transversal extension the assessment of the number of delta rays escaping the detector is complex since the delta rays produced before and after traversing the sensitive volume have a higher chance of being detected by the peripheral part of the detector. Assuming that the particles have a constant LET near the detector, which is a condition generally satisfied for high-energy beam, part of the δ rays that escape from the front and the back of the detector are compensated by the δ rays with identical characteristics entering the detector and generated before and after traversing it. The difference in time of the signal from these δ rays is of the order of one nanosecond and so the contribution to the pulse amplitude is indistinguishable from the contribution of δ rays generated within the detector. Consequently, the sensitive volume for δ rays (intended in the 'dynamic' way just described) is extended by the transversal size of the detector. The rigorous evaluation of the spectral consequences should be based on Monte Carlo simulations. This analysis was performed in some specific cases in a recent study based on GEANT-4 simulations [Solevi et al., 2015].

### 3.7.4. Non-negligible variance of energy loss straggling

The energy loss straggling corresponds to the fluctuation of energy loss by the charge particle crossing the detector and is due to the two stochastic processes, the probability that a collision takes place and the energy lost in an individual collision. The relative variance $V_S$ linked to the uncertainty of these processes can be approximated to a function studied by Kellerer [Kellerer et al., 1975], which can be written emphasizing the dependence on the particles energy, $E$, and mass, $M$:

$$V_S(E, M) = \delta_2/\bar{\epsilon} \cong \frac{E'_{max}}{2\bar{\epsilon}} \Big/ \ln\frac{E'_{max}}{I} \cong \frac{2m_e \cdot E}{M \cdot \bar{\epsilon}} \Big/ \ln\frac{4m_e \cdot E}{M \cdot I} \qquad (28)$$

where $\delta_2$ is the second moment of the δ-ray energy distribution, $\bar{\epsilon}$ is the average energy imparted to the detector by single particles, $I$ is the ionization potential for the target material, $E'_{max}$ is the maximum energy of the δ rays evaluated considering elastic ion-electron collision in the non-relativistic approximation, $m_e$ is the electron mass.

At the right hand side of Eq. 28 the particle energy $E$ appears two times as the part of the ratio, E/M. This ratio is proportional to the energy per nucleon of the carbon ion. It is reasonable to assume that the energy per nucleon of the fragment does not exceed the energy per nucleon of the primary particle and, therefore the variance $V_S$ is valid –or exceeding the actual value– also in case of fragmentation.

Limiting the analysis to the primary carbon ion, if the LET value $L$ is constant within the detector, then $\bar{\epsilon} = L \cdot t$, with $t$ thickness of the detector and, therefore $V_S$ is inversely proportional to the in detector thickness. This observation is used to assess the border of region *IV* in the Fig. 1 considering that $V_S$ should not exceed variance of the chord length distribution, which for spherical sites, correspond to 1/8. The thickness $t'$, of the detectors in which the energy-loss straggling variance equals 1/8 can be expressed as function of the particle energy $E$ by the following:

$$t' \cong \frac{16m_e \cdot E}{M \cdot L} \Big/ \ln\frac{4m_e E}{M \cdot I} \qquad (29)$$

To obtain a *t'* for non-spherical shapes, Eq. 28 should be compared to the variance of chord length for the specific shape. As shown in section 3.2, in slab detectors the variance of the chord length is negligible and therefore the comparison is unworkable. An option is to keep the comparison with the chord length of the sphere also for non-spherical shapes or to assess the threshold of the energy loss straggling from other experimental conditions.

As described by Kellerer [Kellerer, 1970], once the energy-loss straggling is evaluated, is it possible to eliminate its influence to the experimental spectra of lineal energy.

### 3.7.5. Strait line approximation and spectral edges

The accurate assessment of the edges depends in on the trajectory of the primary particle and on the site size. In section 3.1 the assumption for ion-beam therapy was that the particles maintain the same direction and that their paths can be represented by straight parallel lines. In a macroscopic view this condition is confirmed by routine dosimetric controls and it is the base which justifies the use of ion beams to irradiate deep-seated tumors: the so-called pencil beam enters the body with a FWHM in the two transversal directions of few millimeters and is only slightly enlarged after crossing 30 cm of water. This is evidence that the angular distribution on the particle tracks is maintained within angles of the order of the milliradian, with negligible increase of the chord length.

In microscopic view the detour factor is a useful numerical indication which accounts for the trajectory deflection from the axis of the beam. The detour factor is defined as the ratio between the projected range and the continuous-slowing-down approximation (CSDA) ranges. It is always equal to or lower than 1. The CSDA range is a good approximation of the average path length of the primary particles and it corresponds to the integral with respect to the energy of the reciprocal of the total stopping power. The trajectory deflection is noticeable only at the very end of the path. The detour factor in water is close to unity to energies as low as 1 MeV for both, protons (0.99) and carbon ions (0.998). Large deflections of particles have a low probability and negligible effect on the spectrum. For carbon ions, the energy at which the edge is formed is largely above this energy value and this guarantees that the edge is not noticeably affected by particle detours. The detour factor decreases below 90% only for particle energies of few kiloelectronvolts and residual range of less than 2 µm so that the increase of particle path (the reciprocal of the detour factor) is marginal and affecting only the lowest part of the spectrum.

The track of protons in water in the range of energies between 1 MeV and 250 MeV are increased in length by less than 1% on average and therefore the tracks can still be considered as strait and parallel to the beam axis and the distortion can be neglected. At 80 keV, which is the energy of the proton that form the proton edge, the detour factor in water is 0.89 and the projected range is 1.36 µm. The path of the particle in the detector increases because of the detour as well as the probability of a particle to stop inside it. This is resulting in spectral distortions, which affect the spectra at the higher value of lineal energy and the edge in the spectrum.

Similar conclusions for carbon ions and protons are obtained for all typical microdosimetry materials, propane, tissue-equivalent gasses, silicon, and diamond.

### 3.7.6. Edge formation and range factors

Using lookup tables as reference, it is possible to represent the electron stopping power as a function of the projected range, as it is shown in the example of Fig. 4 for carbon ions in graphite at density 3.52 g·cm$^{-3}$.

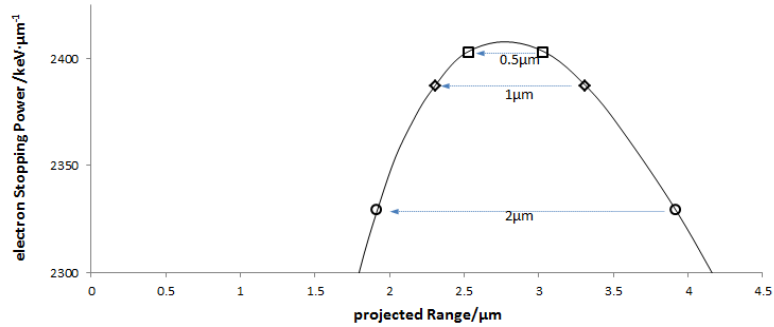

**Figure 4**. Electron stopping power for graphite at density 3.52 g·cm$^{-3}$ as function of the projected range, provided by tabular data of SRIM. The distances of the pair of marker (circles, diamonds, and squares) coincide with the detector thicknesses of 2.0μm, 1.0μm, and 0.5μm, respectively.

In this example it is evident how the condition (i) formulated in section 3.1 is fulfilled in the range region only for the detectors of the smallest size. The maximum electron stopping power in a non-dimensionless detector is a value that depends on the thickness crossed by the particles in the detector. As displayed in Fig. 4 considering the CSDA condition, the maximum energy in electronic collisions is imparted by the ideal particle that enters the sensitive volumes at the right side of the peak and exits, with the same electron stopping power value, at the left side of the peak, crossing a length that corresponds to the detector thickness. The thicker the detector the larger the variation of electron stopping power to take into account. The edges are formed by particles entering with small differences in energy from the ideal particle. For each of the three thicknesses 2.0 μm, 1.0 μm, and 0.5 μm represented in Fig. 4, the average electron stopping power is 2386 keV·μm$^{-1}$, 2401 keV·μm$^{-1}$, and 2406 keV·μm$^{-1}$, respectively. These values should be used as references for the edge values in the three cases.

## 4. Method application and discussion

In this section, the methods described above for the conversion of the spectra to different detector materials (section 3.6.2) and to shapes (section 3.5.1) are applied in sequence using as initial spectrum, a distribution of energy imparted to a 3.1-μm-thick graphite slab representing a simplified detector. The purpose of this example is to show that results that are in agreement with the values foreseen by the theory and indicated in Eq. 13 and Eq. A.7 in Appendix A. The microdosimetric distribution of a graphite slab is obtained using TRIM simulations [Ziegler et al., 2010] and considering a 744 MeV carbon-ion beam crossing 10.9 millimeters of water (which corresponds to the mean range) before reaching the graphite slab. The beam at the entrance is unidirectional, all particles have the same energy, the slab is indefinite extended transversally, and, at density 1 g·cm$^{-3}$, its thickness is 7 μm. A total of $6 \cdot 10^4$ particles is considered, which is a good representation of the number of events that would be collected in clinical irradiation conditions given the area of the detectors and the clinical daily values of dose delivered. In these condition the energy loss straggling is negligible. The continuous line in Fig. 5 represents the microdosimetric spectrum in graphite.

It must be pointed out that the condition (i) indicated in section 3.1 is not satisfied for some of the particles represented in the rightmost part of the spectrum, close to the edge region, and therefore the spectra conversion methods, which assume that the conditions of the method of LET analysis are satisfied, may be imprecise in that area. There are two possible distortions, first on the assessment of the lineal energy value of the edge, which can be corrected as indicated on section 3.7.6, and second on the amplitude of the spectrum in the edge region. These second point can be quantified basing of the Eq. 27 and on the discussion in section 3.7.2. In this example the decreasing of the spectra amplitude in the region of the edges is approximately 4%.

*4.1. Conversion from graphite slab to water slab*

The microdosimetric spectrum converted from graphite slab to water slab with is obtained applying the procedure for the conversion of material described in section 3.5.1 and displayed in Fig. 5 as dashed line. The values of $\bar{y}_F$, $\bar{y}_D$, and $y_{edge}$ for the slab of graphite and water are shown in Table 2.

The ratios of the values $\bar{y}_F$, $\bar{y}_D$, and $y_{edge}$ obtained for slab in water and in graphite are 1.19, 1.22, and 1.37, respectively. If the rescaling was based on a single factor, the three ratios would coincide and the new spectrum would be represented with the same profile of the original simply drifted in a different position. This example, showing a difference of 15% between those ratios, can be seen an indication of the inaccuracies arising from an oversimplified method of material conversion based on a single parameter.

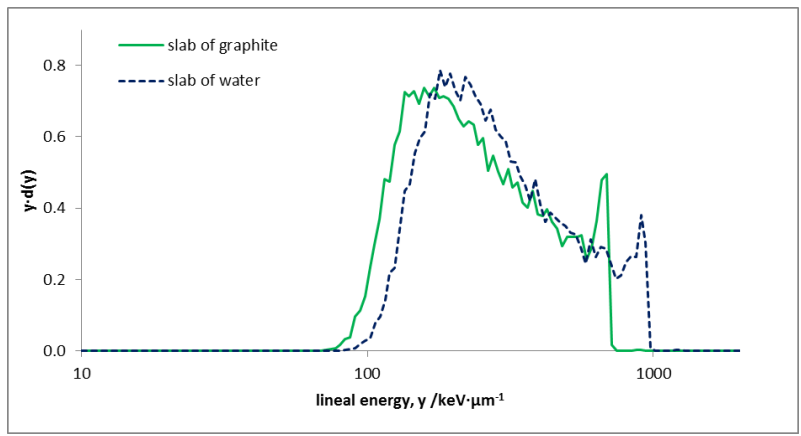

**Figure 5**. Continuous green line: simulated microdosimetric spectrum obtained from 744 MeV carbon-ion beam reaching a 3.1-µm-thick graphite detector after crossing 10.9 mm of water; dashed blue line: microdosimetric spectrum obtained transforming the spectrum in graphite to the spectrum in water.

*4.2. Conversion from slab to sphere and cylinder*

The discrete density spectrum in frequency obtained in water for a slab detector is converted using the algorithm described by Eq. 21 and the cumulative distributions (A.1) and (A.3) to obtain the distribution of lineal energy of spheres and cylinders in water.

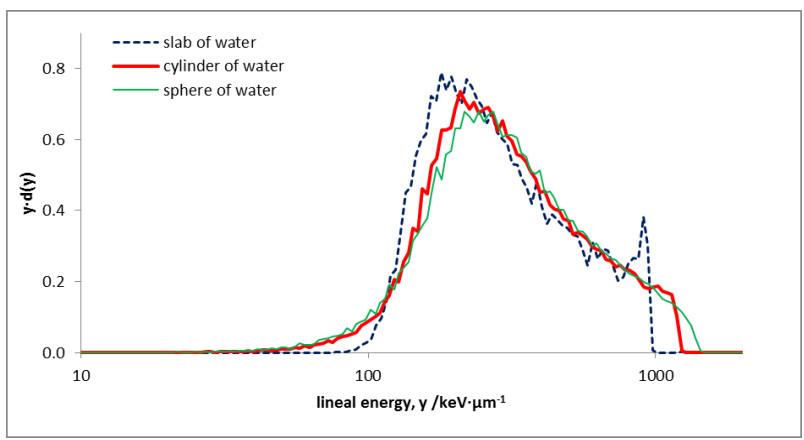

**Figure 6.** Microdosimetric spectra obtained transforming the spectra of lineal energy estimated for the water slab (dashed blue line, as in Fig. 5fig to the spectra that would be obtained with water sphere (green thin line) and a water cylinder (red line thick line).

Table 2 reports also the values of $\bar{y}_F$, $\bar{y}_D$, and $y_{edge}$ evaluated from the spectra displaced in Fig. 6. As foreseen, frequency-mean lineal-energies, $\bar{y}_F$, estimated from the three spectra on water (slab, sphere, and cylinder) coincide in value, 248 keV·µm$^{-1}$. Compared to the value of dose-mean lineal energy $\bar{y}_D$ obtained for the water slab, the value for the sphere is 1.125 times larger and, for the cylinder is 1.081 times larger as predicted by Eq. 13 and Eq. A7. Compared to the edge value of the slab, the value of the cylinder is 1.27 times larger as foreseen in Tab. 1. The edge values in the sphere 1.48 times larger than in the slab, close but not coincident with the value 1.5 reported in Tab. 1 and this small discrepancy is probably due to the intrinsic uncertainties of the method to assess the edge values and to the higher sensitivity of the sphere to the different electron stopping power of water and graphite.

In conclusion, the values estimated for the converted spectra obtained applying the methods introduced in the present study are well in agreement with the theoretical values.

**Table 2.** Estimated values of $\bar{y}_F$, $\bar{y}_D$, and $y_{edge}$ for the four spectra of the example

|  | $\bar{y}_F$ /keV·µm$^{-1}$ | $\bar{y}_D$ /keV·µm$^{-1}$ | $y_{edge}$ /keV·µm$^{-1}$ |
|---|---|---|---|
| Slab of graphite | 207.6 | 273.8 | 699.0 |
| Slab of water | 247.8 | 333.5 | 954.7 |
| Sphere of water | 248.4 | 375.2 | 1415.1 |
| Cylinder of water | 248.2 | 360.5 | 1217.2 |

## 5. Conclusions

This paper studied the characterization the radiation fields of the therapeutic ion beams in terms of radiation quality. The focus is, in particular, on the core of the beam, which is mainly constituted by primary ions with identical directions.

In literature, the radiation quality is specified in terms of LET or in terms of lineal energy. One goal of this study was to pointing out, for the interval of energies typical of the ion beam therapy, differences and similarities of the two specifications looking at the two microdosimetric outcomes, distributions and mean values.

Starting from the simplifications which are at the base of the method of LET analysis and adding the additional condition that particle tracks are essentially unidirectional and parallel, the study identifies as the correlations LET/lineal energy are rather simple and can extend to a large part of the ion-beam therapy energies.

Adjusting the thickness of the detectors considering the beam energy, the lineal energy spectra of spheres and cylinders are expressed as functions of the LET distributions and the mean values of LET and lineal energies are linked by multiplicative factors. In the case of slab detectors the correlation between LET and lineal energy distributions becomes an identity and the multiplicative factor of the mean values is 1.

For energies above 1 MeV/u and below 80 MeV/u, of both, protons and carbon-ions, the simple correlation between LET and lineal energy described above is fulfilled choosing detector sizes (at density 1 g·cm$^{-3}$) between 1µm and 10 µm, which is the order of magnitude of the mammalian cell, the usual reference size of microdosimeters. So, detectors which follow the common definition of 'microdosimeters' are well representing LET parameters.

For energies exceeding 80 MeV/u the size of the detectors which guarantee the correlation between LET and linear energy exceed 10 µm and therefore cannot resemble the cell nucleus size. Those detectors, which cannot strictly be named as microdosimeters, are anyway capable of characterizing univocally the radiation quality in terms of LET.

At energies below 1 MeV/u, which correspond for proton and carbon-ions residual energies in water of 27 μm and 17 μm, respectively, the discrepancies between LET and lineal energy start to be evident.

In this regard, it is important to consider here the different radiation quality assessments in the framework of the clinical use.

When microdosimeters are used to provide information on the radiation quality during the therapy routine, for instance testing the treatment plans in a water phantom, the distortions due to the low-energy particles is marginal due essentially to two factors. First, the role of the range straggling which assures that, until the density of particles per unit of volume is statistically significant, the number of low-energy particles which distort the spectrum, is considerably lower than the number of particles with higher energies. Second, the precision of the positioning procedure is not better than few tenths of millimeters. In that case the uncertainties on the microdosimetric spectra due to the beam position will exceed, by 2 to 3 orders of magnitude, the uncertainties due to the range of low energy particle.

On the other hand, when the microdosimeters are used to characterize the beam for instance to benchmark Monte Carlo simulation studies or to provide information for the creation of a beam model to be used in treatment planning, the variance increase due to low-energy particles must be accounted.

The simple correlation between LET and lineal energy provides, as byproduct, a way to transform the lineal energy spectra collected with one detector to the spectra that would be detected with another detector of different shape and material. The key element is the identity between the LET distributions and the lineal energy spectra collected with ideal slabs. The methodology for the transformation of the experimental spectra is described in an example where the spectrum obtained for slab detector of graphite is converted to the spectrum that would have been collected by detectors of water in the three shapes of slab, sphere, and cylinder.

The methodological observations made in this study serve as a common base for future experimental investigation. The relevance of the method should be tested exposing detectors made of different materials and of different shapes (including one slab) to the same irradiation, and comparing the experimental lineal energy spectra with those obtained from the spectra transformation described here.

The role of the fragments and the potential distortion of the lineal energy spectra should be further investigated experimentally. The method described can also be used to put in evidence the possible discrepancies between the nominal shapes of the sensitive volumes and the actual ones.

# APPENDIX A

## A. Microdosimetric and chord-lengths parameters in spherical and cylindrical sites

Focusing on spherical and cylindrical shapes of the detector sensitive volume, this Appendix reports the cumulative distribution and the density distributions of chord lengths under the assumptions of a negligible variation of the LET while the particle crosses the detector, a negligible fraction of secondary electrons escaping the sensitive volumes, and of straight and parallel particle tracks in the detector. These distributions are used to calculate the first ant the second moments of the chord length and the correlations between lineal energy quantities and their corresponding LET quantities.

### A.1. Cumulative and density distributions

*Spherical detector.* For uniform and unidirectional ion radiation that hits the sphere, the cumulative distribution, $G(\ell)$, represents the probability that the a chord is equal to or less than $\ell$. Using simple geometric considerations $G(\ell)$ is represented as follows:

$$G(\ell) = \begin{cases} \dfrac{\ell^2}{d^2}, & 0 \leq \ell < d \\ 1, & \ell \geq d \end{cases} \quad (A.1)$$

where $d$ is the diameter of the sphere. The probability that a particle crosses the sensitive volume with a chord within the interval $(\ell, \ell + d\ell)$, is expresses by the probability density, $g(\ell)$, which is the derivative of $G(\ell)$ respect to $\ell$:

$$g(\ell) = \frac{dG(\ell)}{d\ell} = \begin{cases} \dfrac{2\ell}{d^2}, & 0 \leq \ell < d \\ 0, & \ell \geq d \end{cases} \quad (A.2)$$

Due to the symmetry of the spherical shape, the functions (A.1) and (A.2) are the same also in the case of isotropic beams.

*Cylindrical detector.* For a uniform and unidirectional irradiation that hits the cylinder perpendicularly to the plane which contains the cylinder axis, the cumulative distribution $C(\ell)$ is expressed as:

$$C(\ell) = \begin{cases} 1 - \sqrt{1 - \dfrac{\ell^2}{d^2}}, & 0 \leq \ell < d \\ 1, & \ell \geq d \end{cases} \quad (A.3)$$

The probability density, $c(\ell)$, derivative of $C(\ell)$ respect to $\ell$ is:

$$c(\ell) = \frac{dC(\ell)}{d\ell} = \begin{cases} \dfrac{\ell}{d^2 \sqrt{1 - \dfrac{\ell^2}{d^2}}}, & 0 \leq \ell < d \\ 0, & \ell \geq d \end{cases} \quad (A.4)$$

where $d$ is the diameter of the cylinder base. These results are independent on the height of the cylinder and so are valid for any cylinder elongations (which is a term used to indicate the ratio between the height and the diameter of the cylinder).

### A.2. Mean chord $\bar{\ell}$ and maximum chord $\ell_{max}$

For unidirectional irradiation directed to a sphere or perpendicular to a cylinder axis, $\ell_{max}$ corresponds to the diameter, $d$ and therefore is results:

in spheres
$$\bar{\ell} = \int_0^\infty \ell \cdot g(\ell)\, d\ell = \int_0^d \frac{2\ell^2}{d^2}\, d\ell = \frac{2}{3}d \cong 0.667 \cdot \ell_{max}$$

in cylinders
$$\bar{\ell} = \int_0^\infty \ell \cdot c(\ell)\, d\ell = \int_0^d \frac{\ell^2}{d^2\sqrt{1-\frac{\ell^2}{d^2}}}\, d\ell = d\frac{\pi}{4} \cong 0.785 \cdot \ell_{max}$$

(A.5)

### A.3. Mean lineal energy values, $\bar{y}_F$ and $\bar{y}_D$,

From Eq. 7, $\overline{\ell^2}$, for the spherical and the cylindrical shapes, becomes:

in spheres
$$\overline{\ell^2} = \int_0^\infty \ell^2 \cdot g(\ell)\, d\ell = \int_0^d \frac{2\ell^3}{d^2}\, d\ell = = \frac{1}{2}\frac{\ell^4}{d^2}\bigg|_0^d = \frac{1}{2}d^2$$

in cylinders
$$\overline{\ell^2} = \int_0^\infty \ell^2 \cdot c(\ell)\, d\ell = \int_0^d \frac{\ell^3}{d^2\sqrt{1-\frac{\ell^2}{d^2}}}\, dl = -\frac{1}{3}(2d^2 + \ell^2)\sqrt{1-\frac{\ell^2}{d^2}}\bigg|_0^d = \frac{2}{3}d^2$$

(A.6)

The dose-mean lineal energy can be represented in terms of dose-mean LET using Eq. 12:

in spheres
$$\bar{y}_D = \bar{L}_D \cdot \frac{1}{2}d^2 \frac{9}{4 \cdot d^2} = \frac{9}{8} \cdot \bar{L}_D = 1.125 \cdot \bar{L}_D$$

in cylinders
$$\bar{y}_D = \bar{L}_D \cdot \frac{2}{3}d^2 \frac{16}{d^2 \cdot \pi^2} = \frac{32}{3\pi^2} \cdot \bar{L}_D \cong 1.081 \cdot \bar{L}_D$$

(A.7)